\documentclass[12pt,fleqn]{article}

\usepackage[latin1]{inputenc}

\usepackage{latexsym,amsmath,amssymb,amscd,amsfonts,amsthm}

\def \Z {\mathbb Z}
\def \R {\mathbb R}
\def \C {\mathbb C}

\def \N {\mathbb N}

\theoremstyle{plain}
\newtheorem{defi}{Definition}
\newtheorem{thm}{Theorem}
\newtheorem{prop}{Proposition}

\newtheorem{lem}{Lemma}

\begin{document}

\title{Positivity of Lyapunov exponents for Anderson-type models on two coupled strings}
\author{Hakim Boumaza\\
Institut de Math\'ematique de Jussieu\\ Universit\'e Paris 7 Denis
Diderot \\ 2 place Jussieu\\ 75251 Paris, France\\ boumaza@math.jussieu.fr\\
\and G\"unter Stolz\\ Department of Mathematics
 CH 452\\ University of Alabama at Birmingham\\ 1300 University
 Boulevard\\
Birmingham, Al 35294-1170, U.S.A.\\ stolz@math.uab.edu }
\date{  }
\maketitle

\abstract{We study two models of Anderson-type random operators on
two deterministically coupled continuous strings. Each model is
associated with independent, identically distributed four-by-four
symplectic transfer matrices, which describe the asymptotics of
solutions. In each case we use a criterion by Gol'dsheid and
Margulis (i.e.\ Zariski denseness of the group generated by the
transfer matrices in the group of symplectic matrices) to prove
positivity of both leading Lyapunov exponents for most energies. In
each case this implies almost sure absence of absolutely continuous
spectrum (at all energies in the first model and for sufficiently
large energies in the second model). The methods used allow for
singularly distributed random parameters, including Bernoulli
distributions.}

\section{Introduction}

Localization for one-dimensional Anderson models is well understood,
while important physical conjectures remain open in dimension $d\ge
2$. In particular, there is no proof yet of the physical conjecture
that, as for $d=1$, localization (in spectral or dynamical sense)
holds at all energies and arbitrary disorder for $d=2$. It is
physically even more convincing that localization should hold for
Anderson models on strips, which should behave like one-dimensional
models.

In fact, Anderson localization has been established rigorously for
discrete strips of arbitrary width in \cite{klein}, for related work
see also \cite{glaffig}. However, an interesting open problem is to
understand the localization properties of Anderson-models on
continuum strips. Consider, for example, the operator
\begin{equation} \label{eq:contstrip}
-\Delta + \sum_{n\in \Z} \omega_n f(x-n,y)
\end{equation}
on $L^2(\R \times [0,1])$ with, say, Dirichlet boundary conditions
on $\R\times\{0\}$ and $\R\times\{1\}$, i.i.d.\ random couplings
$\omega_n$ and a single site potential $f$ supported in $[0,1]\times
[0,1]$. Under weak additional assumptions on $f$ and the
distribution of the $\omega_n$, this operator, describing a
physically one-dimensional disordered system, should be localized at
{\it all} energies. But, with the exception of the easily separable
case of $y$-independent $f$, this question is open. Technically, the
main problem arising is that, while physically one-dimensional, the
model is mathematically multi-dimensional in the sense that the
underlying PDE can not be easily reduced to ODEs. Thus the rich
array of tools for ODEs (coming mostly from dynamical systems) isn't
available. PDE methods like multiscale analysis will show
localization at the bottom of the spectrum (e.g.\ by adapting the
general approach described in \cite{stollmann}), but can't fully
grasp the consequences of physical one-dimensionality.

Thus, one reason for writing this note is to promote the further
study of Anderson models on continuum strips.

Concretely, we take a rather modest step in this direction by
studying two particular models of Anderson-type random operators on
a semi-discretized strip, i.e.\ models of the form
\begin{equation} \label{eq:semidiscrete}
-\frac{d^2}{dx^2} + \sum_{n\in\Z} V(x-n,\omega_n)
\end{equation}
acting on vector-valued functions in $L^2(\R,\C^N)$ for some
positive integer $N$. Here $V(\cdot,\omega)$ is a compactly
supported, $N\times N$-symmetric-matrix-valued random potential, and
$\omega = (\omega_n)_{n\in\Z}$ is a (generally vector-valued)
sequence of i.i.d.\ random variables, to be specified more
explicitly in the models below.

Compared to (\ref{eq:contstrip}), model (\ref{eq:semidiscrete}) is
discretized in the $y$-direction, remaining continuous in the
$x$-direction. This allows to use ODE methods, in particular
transfer matrices and the theory of Lyapunov exponents of products
of independent random matrices. Here the transfer matrices are
$2N$-dimensional and symplectic, leading to $N$ pairs of Lyapunov
exponents $\gamma_1(E) \ge \ldots \ge \gamma_N(E) \ge 0 \ge
-\gamma_N(E) \ge \ldots \ge -\gamma_1(E)$. Kotani-theory for such
operators was developed in \cite{kotanis}. For a non-deterministic
random potential, in particular for model (\ref{eq:semidiscrete}),
the general theory of \cite{kotanis} implies that $\gamma_1(E)>0$
for almost every $E\in\R$. However, for Anderson-type models one
expects that all of the first $N$ Lyapunov exponents are positive
for most energies. This incompleteness of Kotani-theory on the strip
is also pointed out as Problem~3 of the recent review of Kotani
theory in \cite{damanik}.

An abstract criterion for the latter in terms of the groups
generated by the random transfer matrices has been provided by
Gol'dsheid and Margulis \cite{goldsheid}. It is exactly this
criterion which allowed to prove Anderson localization for discrete
strips \cite{klein}.

To the best of our knowledge, our results below are the first
applications of the Gol'dsheid-Margulis criterion to continuum
models. We depend on very explicit calculations and can so far only
handle the case of two coupled strings, i.e.\ $N=2$. Our first model
involves random point interactions, where we can show
$\gamma_2(E)>\gamma_1(E)>0$ for all but an explicitly characterized
discrete set of exceptional energies (Section~\ref{sec:pointint}).
For our second model, two deterministically coupled single string
Anderson models, we get in Section~\ref{sec:anderson} that
$\gamma_2(E) > \gamma_1(E) > 0$ for all but a countable set of
energies $E>2$. As explained at the end of Section~\ref{subsec41},
the latter is a technical restriction and we expect the same to hold
for energies less than $2$.

For both models we conclude the absence of absolutely continuous
spectrum as a consequence of Kotani theory. Discreteness of the set
of exceptional energies, established here for the point interaction
model, should imply that the spectrum is almost surely pure point.
This should follow by extending existing methods, e.g.\
\cite{CKM,klein,stolz2}, but we leave this to a future work.

We start in Section~\ref{sec:separability} with a discussion of the
necessary background on products of i.i.d.\ symplectic matrices and,
in particular, with a statement of the Gol'dsheid-Margulis
criterion.

We mention that quite different methods, going back to the works
\cite{Simon/Spencer} and \cite{KMP}, have been used to prove
localization properties for random operators on strips in
\cite{KMPV}. While \cite{KMPV} only considers discrete strips, the
methods used have potential to be applicable to the continuum strip
model (\ref{eq:contstrip}). One difference between these methods and
the ones used here is that we have aimed at handling singular
distributions of the random parameters, in particular Bernoulli
distributions. This excludes the use of spectral averaging
techniques, which are quite central to the approach in \cite{KMPV}.

The examples studied here are of a very special nature and we hope
to get further reaching results in the future. Still, our simple
examples should be of some interest from the point of view of
exceptional energies where one or several Lyapunov exponents vanish.

It seems that larger $N$ will lead to richer sets of exceptional
energies, as might be expected physically due to the added (at least
partial) transversal degree of freedom of a particle in a strip. Our
examples show that the discrete strip is somewhat untypical in
having no exceptional energies. It has been shown (for models with
$N=1$) that the existence of exceptional energies leads to weaker
localization properties or, more precisely, stronger transport
\cite{JSS,DLS}. A further study of the models proposed here for
larger $N$, with the possibility of observing how this weakens
localization effects, would be quite interesting.

\vskip 3mm

\noindent {\bf Acknowledgements:} This work was partially supported
through NSF-grant DMS-0245210. G.~S.\ is grateful to Anne Boutet de
Monvel and Universit\'e Paris 7 for hospitality and financial
support, which allowed to initiate this collaboration. H.~B.\
enjoyed hospitality at UAB during two visits and received travel
support from Universit\'e Paris 7 and local support from UAB.

\section{Separability of Lyapunov exponents}
\label{sec:separability}

We will first review the main results which allow to prove
simplicity of the Lyapunov spectrum of a sequence of \emph{i.i.d.}
symplectic matrices, and thus, in particular, positivity of the
first $N$ Lyapunov exponents.

\subsection{Lyapunov exponents}

Let $N$ be a positive integer. Let $\mathrm{Sp}_{N}(\R)$ denote the
group of $2N\times 2N$ real symplectic matrices. It is the subgroup
of $\mathrm{GL}_{2N}(\R)$ of matrices $M$ satisfying
$$^tMJM=J,$$
where $J$ is the matrix of order $2N$ defined by $J=\left(
\begin{array}{cc}
0 & -I \\
I & 0
\end{array} \right)$. Here, $I$ is the identity matrix of order $N$.

Recall that for $p\in \{1,\ldots ,N\}$, $\wedge^{p} \R^{N}$ is the
vector space of alternating $p$-linear forms on $(\R^{N})^{*}$. For
$u_{1}, \ldots ,u_{p}$ in $\R^{N}$ and $f_{1}, \ldots ,f_{p}$ in
$(\R^{N})^{*}$, set
$$(u_{1} \wedge \ldots \wedge u_{p})(f_{1},\ldots ,f_{p}) = \det((f_{i}(u_{j}))_{i,j})$$
We call $u_{1} \wedge \ldots \wedge u_{p}$ a decomposable
$p$-vector. We define a basis of $\wedge^{p} \R^{N}$ with those
decomposable $p$-vectors in the following way : if $(u_{1}, \ldots
,u_{N})$ is a basis of $\R^{N}$, $\{ u_{i_{1}} \wedge \ldots \wedge
u_{i_{p}}\ |\ 1\leq i_{1} < \ldots i_{p} \leq N \}$ is a basis of
$\wedge^{p} \R^{N}$. This allows to define all linear operations on
$\wedge^{p} \R^{N}$ on the set of decomposable $p$-vectors.

First, we define a scalar product on $\wedge^{p} \R^{N}$ by the
formula
$$(u_{1} \wedge \ldots \wedge u_{p},v_{1} \wedge \ldots \wedge v_{p})=\det((\langle u_i,v_j \rangle)_{i,j})$$
where $\langle \cdot,\cdot \rangle$ denotes the usual scalar product
on $\R^{N}$. The norm associated with $(\cdot,\cdot)$ will be
denoted by $|| \cdot ||$.

Now we define how an element of the linear group
$\mathrm{GL}_{N}(\R)$ acts on $\wedge^{p} \R^{N}$. If $M\in
\mathrm{GL}_{N}(\R)$, an automorphism $\wedge^{p}M$ of $\wedge^{p}
\R^{N}$ is given by
$$(\wedge^{p}M)(u_{1} \wedge \ldots \wedge u_{p})=Mu_{1} \wedge \ldots \wedge Mu_{p}.$$
We have the property $\wedge^{p}(MN)=(\wedge^{p}M)(\wedge^{p}N)$.
\vskip 3mm \noindent We also introduce the $p$-Lagrangian manifold.
Let $(e_{1},\ldots,e_{2N})$ be the canonical basis of $\R^{2N}$. For
any $p$ in $\{1,\ldots ,N\}$ let $L_{p}$ be the subspace of
$\wedge^{p}\R^{2N}$ spanned by $\{Me_1 \wedge \ldots \wedge Me_p \ |
\ M\in \mathrm{Sp}_{N}(\R) \}$. It is called the $p$-Lagrangian
submanifold of $\R^{2N}$. The projective space $\mathbb{P}(L_p)$ is
the set of isotropic spaces of dimension $p$ in $\R^{2N}$. \vskip
3mm

We can now define the Lyapunov exponents.

\begin{defi}
Let $(A_{n}^{\omega})_{n\in \N}$ be a sequence of i.i.d.\ random
matrices in $\mathrm{Sp}_{N}(\R)$ with $$\mathbb{E}(\log^{+}
||A_{1}^{\omega}||) <\infty.$$ The Lyapunov exponents
$\gamma_{1},\ldots,\gamma_{2N}$ associated with
$(A_{n}^{\omega})_{n\in \N}$ are defined inductively by
$$\sum_{i=1}^{p} \gamma_{i} = \lim_{n \to \infty} \frac{1}{n}
\mathbb{E}(\log ||\wedge^{p} (A_{n}^{\omega}\ldots A_{1}^{\omega})
||).$$
\end{defi}

One has $\gamma_{1} \geq \ldots \geq \gamma_{2N}$ and, due to
symplecticity of the random matrices $(A_{n})_{n\in \N}$, the
symmetry property $\gamma_{2N-i+1}=-\gamma_{i},\ \forall i \in
\{1,\ldots,N\}$ (see \cite{bougerol} p.$89$, Prop $3.2$).

\subsection{A criterion for separability of Lyapunov exponents}

In this section we will follow P.Bougerol and J.Lacroix
\cite{bougerol}. For the definitions of $L_p$-strong irreducibility
and $p$-contractivity we refer to \cite{bougerol}, definitions
$A.IV.3.3$ and $A.IV.1.1$, respectively.

Let $\mu$ be a probability measure on $\mathrm{Sp}_{N}(\R)$. We
denote by $G_{\mu}$ the smallest closed subgroup of
$\mathrm{Sp}_{N}(\R)$ which contains the topological support of
$\mu$, $\mathrm{supp}\ \mu$. \vskip 3mm

Now we can set forth the main result on separability of Lyapunov
exponents, which is a generalization of Furstenberg's theorem to the
case $N>1$.

\begin{prop}\label{main}
Let $(A_{n}^{\omega})_{n\in \N}$ be a sequence of i.i.d.\ random
symplectic matrices of order $2N$ and $p$ be an integer in
$\{1,\ldots ,N\}$. We denote by $\mu$ the common distribution of the
$A_{n}^{\omega}$. Suppose that $G_{\mu}$ is $p$-contracting and
$L_p$-strongly irreducible and that $\mathbb{E}(\log
||A_{1}^{\omega}||)<\infty$. Then the following holds :
\begin{itemize}
\item[(i)] $\gamma_{p} > \gamma_{p+1}$
\item[(ii)] For any non zero $x$ in $L_p$ :
$$\lim_{n\to \infty} \frac{1}{n} \log ||\wedge^{p} A_{n}^{\omega}\ldots A_{1}^{\omega}x||=\sum_{i=1}^{p} \gamma_i\,.$$
\end{itemize}
\end{prop}

This is Proposition~3.4 of \cite{bougerol}, where a proof can be
found. As a corollary we have that if $G_{\mu}$ is $p$-contracting
and $L_p$-strongly irreducible for all $p\in \{1,\ldots ,N\}$ and if
$\mathbb{E}(\log ||A_{1}^{\omega}||)<\infty$, then $\gamma_{1}
> \gamma_{2} > \ldots > \gamma_{N} > 0$ (using the symmetry
property of Lyapunov exponents). \vskip 3mm

For explicit models (that is, explicit $\mu$) it will typically be
quite difficult to check $p$-contractivity and $L_{p}$-strong
irreducibility for all $p$. That is why we will use the
Gol'dsheid-Margulis theory presented in \cite{goldsheid} which gives
us an algebraic argument to verify these assumptions. The idea is
that if the group $G_{\mu}$ is large enough in an algebraic sense
then it is $p$-contractive and $L_{p}$-strongly irreducible for all
$p$.

We recall that the algebraic closure or Zariski closure of a subset
$G$ of an algebraic manifold is the smallest algebraic submanifold
that contains $G$. We denote it by $\mathrm{Cl_{Z}}(G)$. In other
words, if $G$ is a subset of an algebraic manifold, its Zariski
closure $\mathrm{Cl_{Z}}(G)$ is the set of the zeros of polynomials
vanishing on $G$. A subset $G'\subset G$ is said to be Zariski-dense
in $G$ if $\mathrm{Cl_{Z}}(G')=\mathrm{Cl_{Z}}(G)$, \emph{i.e.} each
polynomial vanishing on $G'$ vanishes on $G$. \vskip 3mm More
precisely, from the results of Gol'dsheid and Margulis one easily
gets
\begin{prop}[Gol'dsheid-Margulis criterion] \label{algthm}
If $G_{\mu}$ is Zariski dense in $\mathrm{Sp}_{N}(\R)$, then for all
$p$, $G_{\mu}$ is $p$-contractive and $L_{p}$-strong irreducible.
\end{prop}

\begin{proof}
According to Lemma~$6.2$ and Theorem~$6.3$ on page $57$ of
\cite{goldsheid}, it suffices to prove that the connected component
of the identity of $\mathrm{Sp}_{N}(\R)$ is irreducible in $L_p$ and
that $\mathrm{Sp}_{N}(\R)$ has the $p$-contracting property, for all
$p$. For the $p$-contractivity it suffices to say that
$\mathrm{Sp}_{N}(\R)$ contains an element whose eigenvalues have all
distincts moduli (as an example,
$\mathrm{diag}(2,3,\ldots,N+1,\frac{1}{2},\frac{1}{3},\ldots,\frac{1}{N+1})
\in \mathrm{Sp}_{N}(\R)$) and to use Corollary~$2.2$ in
\cite{bougerol}, p.~82. Next we recall that $\mathrm{Sp}_{N}(\R)$ is
connected and so its connected component of the identity is itself.
And so we have to prove that $\mathrm{Sp}_{N}(\R)$ is irreducible in
$L_p$ for all $p$. This is exactly what is proven in
Proposition~$3.5$ in \cite{bougerol}, p.~91.
\end{proof}

Now we will adopt this algebraic point of view to study two explicit
models.

\section{A model with random point interactions}
\label{sec:pointint}

\subsection{The model}

First, we will study a model of two deterministically coupled
strings with i.i.d.\ point interactions at all integers on both
strings. Formally, this model is given by the random Schrödinger
operator
\begin{equation}\label{model1}
H_{\omega}^P=-\frac{d^{2}}{dx^{2}}+ V_0 +\sum_{n\in \Z} \left(
\begin{array}{cc}
\omega_{1}^{(n)} \delta_{0}(x-n) & 0 \\
0 & \omega_{2}^{(n)} \delta_{0}(x-n)
\end{array}\right)
\end{equation}
acting on $L^{2}(\R,\C^{2})$. Here $V_0$ is the constant-coefficient
multiplication operator by $\left(\begin{matrix} 0 & 1 \\ 1 & 0
\end{matrix}\right)$ and $\delta_{0}$ is the Dirac distribution at
$0$. Also, $\omega^{(n)} = (\omega_1^{(n)}, \omega_2^{(n)})$,
$n\in\Z$, is a sequence of i.i.d.\ $\R^2$-valued random variables
with common distribution $\nu$ on $\R^2$ such that supp$\,\nu
\subset \R^2$ is bounded and not co-linear, i.e.
\begin{equation} \label{notcolinear}
\{x-y: x,y \in \,\mbox{supp}\,\nu\}
\end{equation}
spans $\R^2$. For example, this holds if the components
$\omega_1^{(n)}$ and $\omega_2^{(n)}$ are independent non-trivial
real random variables (i.e.\ each supported on more than one point).

More rigorously,
\begin{equation} \label{model1rig}
H_{\omega}^P = H_{\omega_1} \oplus H_{\omega_2} + V_0
\end{equation}
on $L^2(\R,\C^2) = L^2(\R) \oplus L^2(\R)$, where $H_{\omega_i}$,
$i=1,2$, are operators in $L^2(\R)$ with domain
\begin{eqnarray} \label{mod1domain}
D(H_{\omega_i}) & = & \{ f\in L^2(\R): \;f,f'\:\mbox{are absolutely
continuous on} \: \R\setminus \Z, \: f'' \in L^2(\R), \nonumber \\
& & f \:\mbox{is continuous on}\: \R, \;f'(n^+)=f'(n^-) +
\omega_i^{(n)} f(n) \:\mbox{for all}\:n\in\Z\},
\end{eqnarray}
where existence of the left and right limits $f'(n^-)$ and $f'(n^+)$
at all integers is assumed. On this domain the operator acts by
$H_{\omega_i}f=-f''$. These operators are self-adjoint and bounded
from below, see e.g.\ \cite{albeverioetal}, where boundedness of the
distribution $\nu$ is used. The matrix operator $V_0$ is bounded and
self-adjoint. Thus $H_{\omega}^P$ in (\ref{model1rig}) is
self-adjoint for all $\omega$.

Note here that this model, containing point interactions, is not
covered by the assumptions made in \cite{kotanis}, but that the
proofs easily extend to our setting. Also, \cite{kotanis} considers
$\R$-ergodic systems, while our model is $\Z$-ergodic. However, the
suspension method provided in \cite{kirsch} to extend Kotani-theory
to $\Z$-ergodic operators, also applies to the systems in
\cite{kotanis} and our model. In particular, non-vanishing of all
Lyapunov exponents allows to conclude absence of absolutely
continuous spectrum via an extended version of Theorem~7.2 of
\cite{kotanis}. \vskip 3mm

\vskip 3mm

In order to study the Lyapunov exponents associated with this
operator we need to introduce the sequence of transfer matrices
associated to the equation
\begin{equation}\label{system1}
H_{\omega}^Pu=Eu,\ E\in \R.
\end{equation}
Here, we incorporate the point interactions into the concept of
solution. Thus a function $u=(u_1,u_2):\R\to\C^2$ (not necessarily
square-integrable) is called a solution of (\ref{system1}) if
\begin{equation} \label{freesystem}
-\left( \begin{matrix} u_1 \\ u_2 \end{matrix} \right)'' + V_0
\left(
\begin{matrix} u_1 \\ u_2 \end{matrix} \right) = E \left(
\begin{matrix} u_1 \\ u_2 \end{matrix} \right)
\end{equation}
on $\R\setminus\Z$ and $u$ satisfies the same ``interface
conditions'' as the elements of $D(H_{\omega}^P)$, i.e.\ it is
continuous on $\R$ and
\begin{equation} \label{jump}
u_i'(n^+) =
u_i'(n^-) + \omega_i^{(n)} u_i(n)
\end{equation}
for $i=1,2$ and all $n\in\Z$.

If $u=(u_{1},u_{2})$ is a solution of (\ref{system1}), we define the
transfer matrix $A_{(n,n+1]}^{\omega}(E)$ from $n$ to $n+1$ by the
relation
$$\left( \begin{array}{c}
u_{1}((n+1)^+) \\
u_{2}((n+1)^+) \\
u_{1}'((n+1)^+) \\
u_{2}'((n+1)^+)
\end{array} \right)= A_{(n,n+1]}^{\omega}(E) \left( \begin{array}{c}
u_{1}(n^+) \\
u_{2}(n^+) \\
u_{1}'(n^+) \\
u_{2}'(n^+)
\end{array} \right).$$
Thus we include the effect of the point interaction at $n+1$, but
not at $n$, insuring the usual multiplicative property of transfer
matrices over multiple intervals. The sequence of \emph{i.i.d.}
random matrices $A_{(n,n+1]}^{\omega}(E)$ will determine the
Lyapunov exponents at energy $E$.

By first solving the system (\ref{freesystem}) over $(0,1)$ and then
accounting for the interface condition (\ref{jump}) one can see that
the matrix $A_{(n,n+1]}^{\omega}(E)$ splits into a product of two
matrices:
\begin{equation} \label{matrixsplit}
A_{(n,n+1]}^{\omega}(E)=M(\mbox{diag}(\omega_1^{(n)},\omega_2^{(n)}))
A_{(0,1)}(E)\;.
\end{equation}
Here, for any $2\times 2$-matrix $Q$, we define the $4\times
4$-matrix $M(Q) := \left( \begin{array}{cc} I & 0 \\ Q & I
\end{array} \right)$, where $I$ is the $2\times 2$-unit matrix.
Thus the first factor in (\ref{matrixsplit}) depends only on the
random parameters and
\begin{equation} \label{expform}
A_{(0,1)}(E)=\exp \left(\left( \begin{array}{cccc}
0 & 0 & 1 & 0 \\
0 & 0 & 0 & 1 \\
-E & 1 & 0 & 0 \\
1 & -E & 0 & 0
\end{array}\right) \right)
\end{equation}
depends only on the energy $E$.

The matrices $A_{(n,n+1]}^{\omega}(E)$ are symplectic, which in our
case can be seen directly from their explicit form (but also is part
of the general theory built up in \cite{kotanis}). The distribution
$\mu_E$ of $A_{(0,1]}^{\omega}(E)$ in $\mathrm{Sp}_{2}(\R)$ is given
by
\begin{equation}\label{measuresp}
\mu_E(\Gamma)=\nu (\{\omega^{(0)} \in \R^2:\;
M(\mbox{diag}(\omega_1^{(0)}, \omega_2^{(0)})) A_{(0,1)}(E) \in
\Gamma\})\;.
\end{equation}

The closed group generated by the support of $\mu_E$ is
\begin{equation}\label{groupfd}
G_{\mu_E}=\overline{<M(\mbox{diag}(\omega_1^{(0)}, \omega_2^{(0)}))
A_{(0,1)}(E)|\; \omega^{(0)} \in \,\mbox{supp}\,\nu>}\;.
\end{equation}

The following is our main result for model (\ref{model1}):

\begin{thm}\label{H1thm}
There exists a discrete set $S\subset \R$ such that for all $E \in
\R \setminus S$, $G_{\mu_E}$ is Zariski-dense in
$\mathrm{Sp}_{2}(\R)$. Therefore we have
$\gamma_{1}(E)>\gamma_{2}(E)>0$ for all $E \in \R \setminus S$ and
the operator $H_{\omega}^P$ almost surely has no absolutely
continuous spectrum.
\end{thm}

All we have to prove below is the first statement of
Theorem~\ref{H1thm} about Zariski-denseness. Positivity of Lyapunov
exponents then follows from the results reviewed in
Section~\ref{sec:separability}. As discussed above, Theorem~7.2 of
\cite{kotanis} applies to our model. Thus the essential support of
the a.c.\ spectrum of $H_{\omega}^P$ is contained in the discrete
set $P$, implying that the a.c.\ spectrum is almost surely empty.
\vskip 3mm

The exponential (\ref{expform}) will have different forms for $E>1$,
$E\in (-1,1)$ and $E<-1$. Below we will consider the case $E>1$ in
detail and then briefly discuss the necessary changes for the other
cases. We don't discuss the energies $E=\pm 1$, as we can include
them in the discrete set $S$.

\subsection{Proof of Theorem~\ref{H1thm} for $E>1$}

To study the group $G_{\mu_E}$ we begin by giving an explicit
expression for the transfer matrices. To do this we have to compute
the exponential defining $A_{(0,1)}(E)$. We assume now that $E>1$.
\vskip 3mm \noindent We begin by diagonalizing the real symmetric
matrix $V_0$ in an orthonormal basis:
$$
\left( \begin{array}{cc}
0 & 1 \\
1 & 0
\end{array} \right)= U \left( \begin{array}{cc}
1 & 0 \\
0 & -1
\end{array} \right) U \;.
$$
Here $U=\frac{1}{\sqrt{2}} \left( \begin{array}{cc}
1 & 1 \\
1 & -1
\end{array} \right)$ is orthogonal as well as symmetric. By computing the successive powers of :
$$\left( \begin{array}{cccc}
0 & 0 & 1 & 0 \\
0 & 0 & 0 & 1 \\
-E & 1 & 0 & 0 \\
1 & -E & 0 & 0
\end{array}\right) $$
with each block expressed in the orthonormal basis defined by $U$
one gets
\begin{equation} \label{Ubasis}
A_{(0,1)}(E)=\left( \begin{array}{cc}
U & 0 \\
0 & U
\end{array} \right) \; R_{\alpha,\beta} \; \left( \begin{array}{cc}
U & 0 \\
0 & U
\end{array} \right)\;,
\end{equation}
where $\alpha=\sqrt{E-1}$, $\beta=\sqrt{E+1}$, and
$$
R_{\alpha,\beta}=\left( \begin{array}{cccc}
\cos(\alpha) & 0 & \frac{1}{\alpha}\sin(\alpha) & 0 \\
0 & \cos(\beta) & 0 & \frac{1}{\beta}\sin(\beta) \\
-\alpha \sin(\alpha) & 0 & \cos(\alpha) & 0 \\
0 & -\beta \sin(\beta) & 0 & \cos(\beta)
\end{array}\right)\;.
$$
\vskip 3mm

Now that we have an explicit form for our transfer matrices let us
explain the strategy for proving the Zariski-denseness of
$G_{\mu_E}$ in $\mathrm{Sp}_{2}(\R)$. As $\mathrm{Sp}_{2}(\R)$ is a
connected Lie group, to show Zariski-denseness it is enough to prove
that the Lie algebra of $\mathrm{Cl_{Z}}(G_{\mu_E})$ is equal to the
Lie algebra of $\mathrm{Sp}_{2}(\R)$. The latter is explicitly given
by
$$
\mathfrak{sp}_{2}(\R) = \{ \left( \begin{array}{cc}
a & b_{1} \\
b_{2} & -^{t}a
\end{array} \right),\ a\in \mathrm{M}_{2}(\R),\ b_{1}\ \mathrm{and}\ b_{2}\ \mathrm{symmetric}
\}\;,
$$
which is of dimension 10. So our strategy will be to prove that the
Lie algebra of $\mathrm{Cl_{Z}}(G_{\mu_E})$, which will be denoted
by $\mathfrak{S}_{2}(E)$, is of dimension 10 and to do that we will
explicitly construct 10 linearly independent elements in this Lie
algebra. First we prove

\begin{lem} \label{MQlemma}
For a two-by-two matrix $Q$ one has $M(Q) \in Cl_Z(G_{\mu_E})$ if
and only if $\left( \begin{array}{cc} 0 & 0 \\ Q & 0 \end{array}
\right) \in \mathfrak{S}_{2}(E)$.
\end{lem}

\begin{proof}
If $\left( \begin{array}{cc} 0 & 0 \\ Q & 0 \end{array} \right) \in
\mathfrak{S}_{2}(E)$, then $M(Q) = \exp{(M(Q)-I)} \in
Cl_Z(G_{\mu_E})$. Conversely, if $M(Q) \in Cl_Z(G_{\mu_E})$,
consider the subgroup $G_Q := \{M(nQ) = M(Q)^n:n\in\Z\}$ of
$Cl_Z(G_{\mu_E})$. It follows that $M(xQ) \in Cl_Z(G_Q)$ for all
$x\in\R$. To see this, let $p$ be a polynomial in $4\times 4$
variables such that $p(A)=0$ for all $A\in G_Q$. Then the polynomial
in one variable $\tilde{p}(x) := p(M(xQ))$ has roots in all integers
and must therefore vanish identically. Thus $p(M(xQ))=0$ for all
$x\in\R$. $M(xQ) \in Cl_Z(G_Q) \subset Gl_Z(G_{\mu_E})$ now follows
from the definition of Zariski closure. Then, by differentiating at
the identity element of $Cl_Z(G_{\mu_E})$, we find $\left(
\begin{array}{cc} 0 & 0 \\ Q & 0 \end{array} \right) \in
\mathfrak{S}_{2}(E)$.
\end{proof}

\vskip 3mm

\begin{proof}[Proof of Theorem~\ref{H1thm} for $E>1$]
\noindent \textbf{Step 1.} By (\ref{matrixsplit}),
\begin{equation} \label{aainv}
A_{(0,1]}^{\tilde{\omega}^{(0)}}(E) A_{(0,1]}^{\omega^{(0)}}(E)^{-1}
= M(\mbox{diag}(\tilde{\omega}_1^{(0)}-\omega_1^{(0)},
\tilde{\omega}_2^{(0)}-\omega_2^{(0)})) \in G_{\mu_E}
\end{equation}
for all $\omega^{(0)}, \tilde{\omega}^{(0)} \in \,\mbox{supp}\,\nu$.
As $\mathfrak{S}_{2}(E)$ is an algebra, Lemma~\ref{MQlemma} and
assumption (\ref{notcolinear}) imply that $\left( \begin{array}{cc}
0 & 0 \\ Q & 0 \end{array} \right) \in \mathfrak{S}_{2}(E)$ for
arbitrary diagonal matrices $Q$. \vskip 3mm

\noindent \textbf{Step 2.} Using Step~1 and Lemma~\ref{MQlemma}
shows that $M(Q) \in Cl_Z(G_{\mu_E})$ for arbitrary diagonal $Q$. In
particular, we conclude
$$ A_{(0,1)}(E) = M(\mbox{diag}(\omega_1^{(0)}, \omega_2^{(0)}))^{-1}
A_{(0,1]}(E) \in Cl_Z(G_{\mu_E}).$$ \vskip 3mm

\noindent \textbf{Step 3.} By a general property of matrix Lie
groups we know that
\begin{equation} \label{algconj}
XMX^{-1} \in \mathfrak{S}_{2}(E)
\end{equation}
whenever $M \in \mathfrak{S}_{2}(E)$ and $X\in G_{\mu_E}$. Thus, by
Steps~1 and 2,
\begin{equation} \label{alphabeta}
\left( \begin{array}{cc}
U & 0 \\
0 & U
\end{array} \right)\,R_{\alpha,\beta}^{l}\,\left( \begin{array}{cc}
0 & 0 \\
UQU & 0
\end{array} \right)\,R_{\alpha,\beta}^{-l}\,\left( \begin{array}{cc}
U & 0 \\
0 & U
\end{array} \right) = A_{(0,1)}(E)^l \left( \begin{array}{cc} 0 & 0 \\
Q & 0 \end{array} \right) A_{(0,1)}(E)^{-l} \in \mathfrak{S}_{2}(E),
\end{equation}
where $Q=\mbox{diag}(\omega_{1}^{(0)}, \omega_{2}^{(0)})$. But we
also have, as $U$ is orthogonal and symmetric,
$$
\mathfrak{S}_{2}(E)=\mathfrak{sp}_{2}(\R) \Leftrightarrow
\tilde{\mathfrak{S}}_{2}(E) := \left(
\begin{array}{cc}
U & 0 \\
0 & U
\end{array} \right)\mathfrak{S}_{2}(E)\left( \begin{array}{cc}
U & 0 \\
0 & U
\end{array} \right)=\mathfrak{sp}_{2}(\R)
$$

Thus we are left with having to show the latter. To this end, we
know from (\ref{alphabeta}) that
\begin{equation} \label{tildealphabeta}
S(l,Q) := R_{\alpha,\beta}^{l}\,\left( \begin{array}{cc}
0 & 0 \\
UQU & 0
\end{array} \right)\,R_{\alpha,\beta}^{-l} \in \tilde{\mathfrak{S}}_{2}(E)
\end{equation}
for all $l\in\Z$ and all four matrices
$Q=\mbox{diag}(\omega_1^{(0)}, \omega_2^{(0)})$. \vskip 3mm

\noindent \textbf{Step 4.} By Step 3, for
$(\omega_{1}^{(0)},\omega_{2}^{(0)})=(1,1)$ and all $l\in\Z$,
\begin{eqnarray*}
\lefteqn{A_{1}(l):= S(l,I) =} \nonumber \\ && \left(
\begin{array}{cccc}
\frac{1}{\alpha} \sin(l\alpha) \cos(l\alpha) & 0 & -\frac{1}{\alpha^{2}}\sin^{2}(l\alpha) & 0 \\
0 & \frac{1}{\beta} \sin(l\beta) \cos(l\beta) & 0 & -\frac{1}{\beta^{2}}\sin^{2}(l\beta) \\
\cos^{2}(l\alpha) & 0 & -\frac{1}{\alpha}\cos(l\alpha)\sin(l\alpha) & 0 \\
0 & \cos^{2}(l\beta) & 0 & -\frac{1}{\beta} \sin(l\beta) \cos(l\beta) \\
\end{array}\right)\in \tilde{\mathfrak{S}}_{2}(E).
\end{eqnarray*}
Also, as $\tilde{\mathfrak{S}}_{2}(E)$ is an algebra, $A_2(l) :=
2S(l,\left( \begin{matrix} 1 & 0 \\ 0 & 0 \end{matrix} \right)) -
S(l,I) \in \tilde{\mathfrak{S}}_{2}(E)$ for all $l\in \Z$. These
matrices take the form
$$
A_{2}(l)=\left( \begin{array}{cccc}
0 & \frac{1}{\alpha} \sin(l\alpha) \cos(l\beta) & 0 & -\frac{1}{\alpha \beta}\sin(l\alpha)\sin(l\beta) \\
\frac{1}{\beta} \sin(l\beta) \cos(l\alpha) & 0 & -\frac{1}{\alpha \beta}\sin(l\alpha)\sin(l\beta) & 0 \\
0 & \cos(l\alpha)\cos(l\beta) & 0 & -\frac{1}{\beta}\cos(l\alpha)\sin(l\beta)\\
\cos(l\alpha)\cos(l\beta) & 0 & -\frac{1}{\alpha} \sin(l\alpha) \cos(l\beta) & 0 \\
\end{array}\right)\in \tilde{\mathfrak{S}}_{2}(E).
$$
\vskip 3mm

\noindent \textbf{Step 5.} We remark that the space generated by the
family $(A_{1}(l))_{l\in \Z}$ is orthogonal to the one generated by
$(A_{2}(l))_{l\in \Z}$ in $\mathfrak{sp}_{2}(\R)$. We can work
independently with each of these two families to find enough
linearly independent matrices in $\tilde{\mathfrak{S}}_{2}(E)$ to
generate a subspace of dimension 10. We begin with the family
$(A_{2}(l))_{l\in \Z}$. We want to prove that for all but a discrete
set of energies $E\in \R$, $A_{2}(0),A_{2}(1),A_{2}(2),A_{2}(3)$ are
linearly independent. Because of the symmetries in the coefficients
of these matrices, their linear independence is equivalent to the
linear independence of the vectors
$$\left( \begin{array}{c}
1 \\
0 \\
0 \\
0
\end{array}\right),\ \left( \begin{array}{c}
\cos(\alpha)\cos(\beta) \\
-\frac{1}{\alpha \beta} \sin(\alpha) \sin(\beta) \\[1mm]
\frac{1}{\alpha} \sin(\alpha) \cos(\beta) \\[1mm]
\frac{1}{\beta} \sin(\beta) \cos(\alpha)
\end{array}\right) ,\ \left( \begin{array}{c}
\cos(2\alpha)\cos(2\beta) \\
-\frac{1}{\alpha \beta} \sin(2\alpha) \sin(2\beta) \\[1mm]
\frac{1}{\alpha} \sin(2\alpha) \cos(2\beta) \\[1mm]
\frac{1}{\beta} \sin(2\beta) \cos(2\alpha)
\end{array}\right),\ \left( \begin{array}{c}
\cos(3\alpha)\cos(3\beta) \\
-\frac{1}{\alpha \beta} \sin(3\alpha) \sin(3\beta) \\[1mm]
\frac{1}{\alpha} \sin(3\alpha) \cos(3\beta) \\[1mm]
\frac{1}{\beta} \sin(3\beta) \cos(3\alpha)
\end{array}\right)\;.$$
The determinant of the matrix generated by those four vectors, best
found with the help of a computer algebra system, is
\begin{equation}\label{det11}
 \frac{4}{\alpha^{2} \beta^{2}}\sin^{2}(\alpha) \sin^{2}(\beta)
 (\cos^{2}(\alpha)-\cos^{2}(\beta))\;.
\end{equation}
This function is real-analytic in $E>1$ with roots not accumulating
at $1$, thus it vanishes only for a discrete set $S_{1}$ of energies
$E>1$. \vskip 3mm

\noindent \textbf{Step 6.} By Step 5 we know that
$(A_{2}(0),A_{2}(1),A_{2}(2),A_{2}(3))$ generate a subspace of
dimension four of $\tilde{\mathfrak{S}}_{2}(E)$ for $E\in (1,\infty)
\setminus S_{1}$, i.e.\ they generate all matrices of the form
$$
\left( \begin{array}{cccc}
0 & a & 0 & d \\
b & 0 & d & 0 \\
0 & c & 0 & -b \\
c & 0 & -a & 0
\end{array} \right)\;,
$$
with $a,b,c,d\in \R$. In particular, for $a=1$, $b=c=d=0$,
$$
B_0 := \left( \begin{array}{cccc}
0 & 1 & 0 & 0 \\
0 & 0 & 0 & 0 \\
0 & 0 & 0 & 0 \\
0 & 0 & -1 & 0
\end{array} \right) \in  \tilde{\mathfrak{S}}_{2}(E)\;.
$$
It follows that $B:= \frac{1}{2}[B_0,A_2(0)] \in
\tilde{\mathfrak{S}}_{2}(E)$, with $[\cdot,\cdot]$ denoting the
matrix commutator bracket. We calculate
$$  B =  \left( \begin{array}{cccc}
0 & 0 & 0 & 0 \\
0 & 0 & 0 & 0 \\
0 & 0 & 0 & 0 \\
0 & 1 & 0 & 0
\end{array} \right)\;.$$
\vskip 3mm

\noindent \textbf{Step 7.} In this step we will prove that
$A_{1}(0),A_{1}(1),A_{1}(2),A_{1}(3),A_{1}(4),B$ are linearly
independent for all but a discrete set of energies. Due to the
symmetries and zeros in matrices of the form $A_1(l)$, it suffices
to show linear independence of the vectors formed by the entries
$(3,1), (4,2), (1,1), (2,2), (1,3), (2,4)$ of these six matrices.
The determinant of the matrix spanned by these six columns is found
to be
\begin{equation}\label{det12}
\frac{64}{\alpha^{3}\beta^{3}}\sin^{3}(\alpha)\sin^{3}(\beta)\cos(\alpha)
\cos(\beta) (\cos^{2}(\alpha)-\cos^{2}(\beta))^{2}\;.
\end{equation}
Similar to Step 5 one argues that this vanishes only for $E$ in
$\mathfrak{sp}_{2}(\R)$ a discrete subset $S_2\subset (1,\infty)$.
Thus, for $E\in (1,+\infty) \setminus S_{2}$,
$A_{1}(0),A_{1}(1),A_{1}(2),A_{1}(3),A_{1}(4),B$ are linearly
independent. \vskip 3mm

\noindent \textbf{Step 8.} First we can see that $S_{1} \subset
S_{2}$. Having complementary sets of non-zero entries, the subspaces
generated by $A_{1}(0)$, $A_{1}(1)$, $A_{1}(2)$, $A_{1}(3)$,
$A_{1}(4),B$ as well as $A_{2}(0)$, $A_{2}(1)$, $A_{2}(2)$,
$A_{2}(3)$ are orthogonal. If $E\in (1,\infty) \setminus S_2$, then
both sets of matrices are linearly independent and contained in
$\tilde{\mathfrak{S}}_{2}(E)$. Thus the latter has at least
dimension 10 and therefore is equal to $\mathfrak{sp}_{2}(\R)$. This
concludes the proof of Theorem~\ref{H1thm} for the case $E>1$.
\end{proof}

\subsection{The cases $E\in (-1,1)$ and $E<-1$}

We now turn to the proof of the Theorem for $E\in (-1,1)$. Here the
expression for the matrix $A_{(0,1)}(E)$ changes slightly. We now
set $\alpha=\sqrt{1-E}$ and, as before, $\beta=\sqrt{E+1}$. Also,
$U$ remains unchanged. But we replace $R_{\alpha,\beta}$ by
$$
\tilde{R}_{\alpha,\beta}=\left( \begin{array}{cccc}
\cosh(\alpha) & 0 & \frac{1}{\alpha}\sinh(\alpha) & 0 \\
0 & \cos(\beta) & 0 & \frac{1}{\beta}\sin(\beta) \\
\alpha \sinh(\alpha) & 0 & \cosh(\alpha) & 0 \\
0 & -\beta \sin(\beta) & 0 & \cos(\beta)
\end{array}\right)\;.
$$

\begin{proof}[Proof of Theorem~\ref{H1thm} for $E\in (-1,1)$]
In fact, we can follow the proof for the first case very closely. We
will briefly comment on the changes. Steps 1 and 2 remain unchanged.
In the Step 3 we replace $R_{\alpha,\beta}$ by
$\tilde{R}_{\alpha,\beta}$ and so in Step 4 we get that for all
$l\in\Z$,
\begin{eqnarray*}
\lefteqn{\tilde{A}_{1}(l):=} \\ && \left( \begin{array}{cccc}
\frac{1}{\alpha} \sinh(l\alpha) \cosh(l\alpha) & 0 & -\frac{1}{\alpha^{2}}\sinh^{2}(l\alpha) & 0 \\
0 & \frac{1}{\beta} \sin(l\beta) \cos(l\beta) & 0 & -\frac{1}{\beta^{2}}\sin^{2}(l\beta) \\
\cosh^{2}(l\alpha) & 0 & -\frac{1}{\alpha}\cosh(l\alpha)\sinh(l\alpha) & 0 \\
0 & \cos^{2}(l\beta) & 0 & -\frac{1}{\beta} \sin(l\beta) \cos(l\beta) \\
\end{array}\right)
\end{eqnarray*}
is in $\mathfrak{S}_{2}(E)$, and
\begin{eqnarray*}
\lefteqn{\tilde{A}_{2}(l):=} \\ && \left( \begin{array}{cccc}
0 & \frac{1}{\alpha} \sinh(l\alpha) \cos(l\beta) & 0 & -\frac{1}{\alpha \beta}\sinh(l\alpha)\sin(l\beta) \\
\frac{1}{\beta} \sin(l\beta) \cosh(l\alpha) & 0 & -\frac{1}{\alpha \beta}\sinh(l\alpha)\sin(l\beta) & 0 \\
0 & \cosh(l\alpha)\cos(l\beta) & 0 & -\frac{1}{\beta}\cosh(l\alpha)\sin(l\beta)\\
\cosh(l\alpha)\cos(l\beta) & 0 & -\frac{1}{\alpha} \sinh(l\alpha) \cos(l\beta) & 0 \\
\end{array}\right)
\end{eqnarray*}
is in $\mathfrak{S}_{2}(E)$. In Step 5 we again get that for all but
a discrete set of energies, $\tilde{A}_{2}(0)$, $\tilde{A}_{2}(1)$,
$\tilde{A}_{2}(2)$, $\tilde{A}_{2}(3)$ are linearly independent. The
determinant set up from the entries in exactly the same way as in
Step 5 above is now
\begin{equation}\label{det21}
\frac{4}{\alpha^{2} \beta^{2}}\sinh^{2}(\alpha) \sin^{2}(\beta)
(\cosh^{2}(\alpha)-\cos^{2}(\beta))\;,
\end{equation}
which vanishes only on a finite set $S_3$ of values $E\in (-1,1)$.

Step 6 remains unchanged except we now get $B\in
\mathfrak{S}_{2}(E)$ for all $E\in (-1,1) \setminus S_{3}$. In
Step~7 we set up a $6\times 6$-matrix from the entries of
$\tilde{A}_{1}(0), \ldots, \tilde{A}_{1}(4), B$ in exactly the same
way as in Step~7 above and find for its determinant
\begin{equation} \label{det22}
\frac{64}{\alpha^{3}\beta^{3}}\sinh^{3}(\alpha)\sin^{3}(\beta)\cosh(\alpha)
\cos(\beta) (\cosh^{2}(\alpha)-\cos^{2}(\beta))^{2}\;.
\end{equation}
The roots of this function are a subset $S_4$ of $(-1,1)$, which
contains $S_3$. As in Step 8 we conclude that for all energies $E\in
(-1,1)\setminus S_{4}$,
$\mathrm{Cl_{Z}}(G_{\mu_E})=\mathrm{Sp}_{2}(\R)$.
\end{proof}

Finally, without providing further details, we note that very
similar changes can be used to cover the remaining case $E<-1$.

\section{Matrix-valued continuum Anderson model}
\label{sec:anderson}

While in our first model the randomness acted through point
interactions on a discrete set, we now turn to a model with more
extensive randomness. We consider two independent continuum Anderson
models on single strings, with the single site potentials given by
characteristic functions of unit intervals, and couple the two
strings with the deterministic off-diagonal matrix $V_0$ already
used above.

\subsection{The model} \label{subsec41}

Let
\begin{equation}\label{model2}
H_{\omega}^A=-\frac{d^{2}}{dx^{2}} + V_0 + \sum_{n\in \Z} \left(
\begin{array}{cc}
\omega_{1}^{(n)} \chi_{[0,1]}(x-n) & 0 \\
0 & \omega_{2}^{(n)} \chi_{[0,1]}(x-n)
\end{array} \right)
\end{equation}
be a random Schr\"odinger operator acting in $L^{2}(\R,\C^{2})$,
where  $\chi_{[0,1]}$ is the characteristic function of the interval
$[0,1]$, $V_0$ is as in the previous model, and
$(\omega_{1}^{(n)})_{n\in\Z}$ and $(\omega_{2}^{(n)})_{n\in\Z}$ are
two sequences of i.i.d.\ random variables (also independent from
each other) with common distribution $\tilde{\nu}$ such that
$\{0,1\} \subset \;\mbox{supp}\,\tilde{\nu}$.

This operator is a bounded perturbation of $(-\frac{d^{2}}{dx^{2}})
\oplus (-\frac{d^{2}}{dx^{2}})$ and thus self-adjoint on the
$\C^2$-valued second order $L^2$-Sobolev space. \vskip 3mm

For this model we have the following result:

\begin{thm}\label{H2thm}
There exists a countable set $\mathcal{C}$ such that for all $E\in
(2,\infty ) \setminus \mathcal{C}$, $\gamma_{1}(E)
> \gamma_{2}(E) >0$. Therefore, $H_{\omega}^A$ has no absolutely
continuous spectrum in the interval $(2,\infty )$.
\end{thm}
\vskip 3mm

The transfer matrices $A_{n,2}^{\omega}(E)$, mapping $(u_1(n),
u_2(n), u_1'(n), u_2'(n))$ to $(u_1(n+1), u_2(n+1), u_1'(n+1),
u_2'(n+1))$ for solutions $u=(u_1,u_2)$ of the equation
$H_{\omega}^Au=Eu$, are i.i.d.\ and symplectic \cite{kotanis}.
Denote the distribution of $A_{0,2}^{\omega}(E)$ in $Sp_2(\R)$ by
$\tilde{\mu}_E$. As before, by $G_{\tilde{\mu}_E}$ we denote the
closed subgroup of $Sp_2(\R)$ generated by supp$\,\tilde{\mu}_E$. As
$\{0,1\} \subset \;\mbox{supp}\,\tilde{\nu}$ we have that
$$
\{A_{0,2}^{(0,0)}(E), A_{0,2}^{(1,0)}(E), A_{0,2}^{(0,1)}(E),
A_{0,2}^{(1,1)}(E)\} \subset G_{\tilde{\mu}_E}\;.
$$
Here we also write $A_{0,2}^{\omega^{(0)}}(E)$ for the transfer
matrices from $0$ to $1$, where $\omega^{(0)} = (\omega_1^{(0)},
\omega_2^{(0)})$. We will denote the Lie algebra of the Zariski
closure $\mathrm{Cl_{Z}}(G_{\tilde{\mu}_E})$ of $G_{\tilde{\mu}_E}$
by $\mathfrak{A}_{2}(E)$.

To give an explicit description of the matrices
$A_{0,2}^{\omega^{(0)}}(E)$ we define,
$$
M_{\omega^{(0)}}=\left( \begin{array}{cc}
\omega_{1}^{(0)} & 1 \\
1 & \omega_{2}^{(0)}
\end{array} \right) = S_{\omega^{(0)}} \left( \begin{array}{cc}
\lambda_{1}^{\omega^{(0)}} & 0 \\
0 & \lambda_{2}^{\omega^{(0)}}
\end{array} \right)S_{\omega^{(0)}}^{-1},
$$
with orthogonal matrices $S_{\omega^{(0)}}$ and the real eigenvalues
$\lambda_{1}^{\omega^{(0)}}\leq \lambda_{2}^{\omega^{(0)}}$ of
$M_{\omega^{(0)}}$. Explicitly, we get
$$
S_{(0,0)}=\frac{1}{\sqrt{2}}\left( \begin{array}{cc}
1 & 1 \\
1 & -1
\end{array} \right),\ \lambda_{1}^{(0,0)}=1,\ \lambda_{2}^{(0,0)}=-1,
$$
$$
S_{(1,1)}=S_{(0,0)},\ \lambda_{1}^{(1,1)}=2,\ \lambda_{2}^{(1,1)}=0,
$$
$$
S_{(1,0)}=\left( \begin{array}{cc}
\frac{2}{\sqrt{10-2\sqrt{5}}} & \frac{2}{\sqrt{10+2\sqrt{5}}} \\[1mm]
\frac{-1+\sqrt{5}}{\sqrt{10-2\sqrt{5}}} &
\frac{-1-\sqrt{5}}{\sqrt{10+2\sqrt{5}}}
\end{array} \right),\ \lambda_{1}^{(1,0)}=\frac{1+\sqrt{5}}{2},\
\lambda_{2}^{(1,0)}=\frac{1-\sqrt{5}}{2}\;.
$$
We don't compute $S_{(0,1)},
\lambda_{1}^{(0,1)},\lambda_{2}^{(0,1)}$ because we will not use
them in the following.

We also introduce the block matrices
$$
R_{\omega^{(0)}}=\left( \begin{array}{cc}
S_{\omega^{(0)}} & 0 \\
0 & S_{\omega^{(0)}}
\end{array} \right)\;.
$$

Let $E>2$ (and thus larger than all eigenvalues of all
$M_{\omega^{(0)}}$). With the abbreviation $r_i =
r_i(E,\omega^{(0)}) := \sqrt{E-\lambda_i^{\omega^{(0)}}}$, $i=1,2$,
the transfer matrices become
\begin{equation} \label{expltransfer}
A_{0,2}^{\omega^{(0)}}(E)=R_{\omega^{(0)}}\left( \begin{array}{cccc}
\cos(r_{1}) & 0 & \frac{1}{r_{1}} \sin(r_{1}) & 0 \\[1mm]
0 & \cos(r_{2}) & 0 & \frac{1}{r_{2}} \sin(r_{2}) \\[1mm]
-r_{1} \sin(r_{1}) & 0 & \cos(r_{1}) & 0 \\[1mm]
0 & -r_{2} \sin(r_{2}) & 0 & \cos(r_{2})
\end{array} \right) R_{\omega^{(0)}}^{-1}\;.
\end{equation}

For $E<2$ one can still write explicit expressions for the transfer
matrices, where some of the sines and cosines are replaced by the
respective hyperbolic functions, depending on the relative location
of $E$ to the various $\lambda_i^{\omega^{(0)}}$. This would lead to
various cases, for each of which the arguments of the following
subsection are not quite as easily adjustable as in the cases of
Section~\ref{sec:pointint}. We therefore left the nature of Lyapunov
exponents for $E\in (-1,2)$ open, even if we fully expect similar
results. As in Section~\ref{sec:pointint} it is seen that the
minimum of the almost sure spectrum of $H_{\omega}^A$ is again $-1$.

\subsection{Proof of Theorem~\ref{H2thm}}

Using the Gol'dsheid-Margulis criterion and the results from
\cite{kotanis} and \cite{kirsch} as in Section~\ref{sec:pointint}
above, it will suffice to prove the following :

\begin{prop}\label{H2zar}
There exists a countable set $\mathcal{C}$ such that for all $E\in
(2,\infty ) \setminus \mathcal{C}$, $G_{\tilde{\mu}_E}$ is
Zariski-dense in $\mathrm{Sp}_{2}(\R)$.
\end{prop}

\begin{proof}
\textbf{Step 1.} We fix $E\in (2,\infty )$. For $\omega^{(0)}=(0,0)$
we have
\begin{equation}\label{A00}
A_{0,2}^{(0,0)}(E)=R_{(0,0)}\left( \begin{array}{cccc}
\cos(\alpha_{1}) & 0 & \frac{1}{\alpha_{1}} \sin(\alpha_{1}) & 0 \\[1mm]
0 & \cos(\alpha_{2}) & 0 & \frac{1}{\alpha_{2}} \sin(\alpha_{2}) \\[1mm]
-\alpha_{1} \sin(\alpha_{1}) & 0 & \cos(\alpha_{1}) & 0 \\[1mm]
0 & -\alpha_{2} \sin(\alpha_{2}) & 0 & \cos(\alpha_{2})
\end{array} \right) R_{(0,0)}^{-1}
\end{equation}
where $\alpha_{1}=\sqrt{E-\lambda_{1}^{(0,0)}}=\sqrt{E-1}$ and
$\alpha_{2}=\sqrt{E-\lambda_{2}^{(0,0)}}=\sqrt{E+1}$.

Let $\mathcal{C}_{1}$ be the set of energies such that
$(2\pi,\alpha_{1},\alpha_{2})$ is a rationally dependent set. It is
easily checked that $\mathcal{C}_{1}$ is countable.

We now assume that $E\in (2,\infty ) \setminus \mathcal{C}_{1}$.
Rational independence of $(2\pi,\alpha_{1},\alpha_{2})$ implies that
there exists a sequence $(n_{k})\in \N^{\N}$, such that
$$
(n_{k}\alpha_{1},n_{k}\alpha_{2})\xrightarrow[k\to \infty]{}
(\frac{\pi}{2},0)
$$
with convergence in $\R^{2}/(2\pi \Z)^{2}$. There also exist
$(m_{k})\in \N^{\N}$, such that
$$
(m_{k}\alpha_{1},m_{k}\alpha_{2})\xrightarrow[k\to \infty]{}
(0,\frac{\pi}{2})\;.
$$

Then, as $G_{\tilde{\mu}_E}$ is closed, we conclude that
\begin{equation}\label{alpha1G}
\left( A_{0,2}^{(0,0)}(E) \right)^{n_{k}}\xrightarrow[k\to \infty]{}
R_{(0,0)} \left( \begin{array}{cccc}
0 & 0 & \frac{1}{\alpha_{1}} & 0 \\
0 & 1 & 0 & 0 \\
-\alpha_{1} & 0 & 0 & 0 \\
0 & 0 & 0 & 1
\end{array} \right)R_{(0,0)}^{-1} \in G_{\tilde{\mu}_E}
\end{equation}
and
\begin{equation}\label{alpha2G}
\left( A_{0,2}^{(0,0)}(E) \right)^{m_{k}}\xrightarrow[k\to \infty]{}
R_{(0,0)} \left( \begin{array}{cccc}
1 & 0 & 0 & 0 \\
0 & 0 & 0 & \frac{1}{\alpha_{2}} \\
0 & 0 & 1 & 0 \\
0 & -\alpha_{2} & 0 & 0
\end{array} \right)R_{(0,0)}^{-1} \in G_{\tilde{\mu}_E}\;.
\end{equation}
\vskip 3mm

For $\omega^{(0)}=(1,1)$ we have
\begin{equation}\label{A11}
A_{0,2}^{(1,1)}(E)=R_{(0,0)}\left( \begin{array}{cccc}
\cos(\beta_{1}) & 0 & \frac{1}{\beta_{1}} \sin(\beta_{1}) & 0 \\[1mm]
0 & \cos(\beta_{2}) & 0 & \frac{1}{\beta_{2}} \sin(\beta_{2}) \\[1mm]
-\beta_{1} \sin(\beta_{1}) & 0 & \cos(\beta_{1}) & 0 \\[1mm]
0 & -\beta_{2} \sin(\beta_{2}) & 0 & \cos(\beta_{2})
\end{array} \right) R_{(0,0)}^{-1}\;,
\end{equation}
where $\beta_{1}=\sqrt{E-\lambda_{1}^{(1,1)}}=\sqrt{E-2}$ and
$\beta_{2}=\sqrt{E-\lambda_{2}^{(1,1)}}=\sqrt{E}$.

Similarly, working with powers of $A_{0,2}^{(1,1)}(E)$, we see that
for $E$ such that $(2\pi,\beta_{1},\beta_{2})$ is rationally
independent (which occurs away from a countable set
$\mathcal{C}_{2}$)
\begin{equation}\label{beta1G}
R_{(0,0)} \left( \begin{array}{cccc}
0 & 0 & \frac{1}{\beta_{1}} & 0 \\
0 & 1 & 0 & 0 \\
-\beta_{1} & 0 & 0 & 0 \\
0 & 0 & 0 & 1
\end{array} \right)R_{(0,0)}^{-1} \in G_{\tilde{\mu}_E}
\end{equation}
and
\begin{equation}\label{beta2G}
R_{(0,0)} \left( \begin{array}{cccc}
1 & 0 & 0 & 0 \\
0 & 0 & 0 & \frac{1}{\beta_{2}} \\
0 & 0 & 1 & 0 \\
0 & -\beta_{2} & 0 & 0
\end{array} \right)R_{(0,0)}^{-1} \in G_{\tilde{\mu}_E}\;.
\end{equation}
\vskip 3mm

\noindent \textbf{Step 2.} Multiplying (\ref{alpha1G}) by the
inverse of (\ref{beta1G}) we get
$$
R_{(0,0)} \left( \begin{array}{cccc}
\frac{\beta_{1}}{\alpha_{1}} & 0 & 0 & 0 \\
0 & 1 & 0 & 0 \\
0 & 0 & \frac{\alpha_{1}}{\beta_{1}} & 0 \\
0 & 0 & 0 & 1
\end{array} \right)R_{(0,0)}^{-1} \in G_{\tilde{\mu}_E}\;.
$$
As $\alpha_{1} > \beta_{1} >0$, by an argument similar to the one
used in the proof of Lemma~\ref{MQlemma}, this implies that for all
$x>0$,
$$
C_{1}(x)=R_{(0,0)} \left( \begin{array}{cccc}
x & 0 & 0 & 0 \\
0 & 1 & 0 & 0 \\
0 & 0 & \frac{1}{x} & 0 \\
0 & 0 & 0 & 1
\end{array} \right)R_{(0,0)}^{-1} \in
\mathrm{Cl_{Z}}(G_{\tilde{\mu}_E})\;.
$$

We remark that $C_{1}(1)=I$. Thus, by differentiating at $I$,
$$
C_{1}:=R_{(0,0)} \left( \begin{array}{cccc}
1 & 0 & 0 & 0 \\
0 & 0 & 0 & 0 \\
0 & 0 & -1 & 0 \\
0 & 0 & 0 & 0
\end{array} \right)R_{(0,0)}^{-1} \in \mathfrak{A}_{2}(E)\;.
$$

In the same way, using (\ref{alpha2G}) and (\ref{beta2G}),
$$
C_{2}:=R_{(0,0)} \left( \begin{array}{cccc}
0 & 0 & 0 & 0 \\
0 & 1 & 0 & 0 \\
0 & 0 & 0 & 0 \\
0 & 0 & 0 & -1
\end{array} \right)R_{(0,0)}^{-1} \in \mathfrak{A}_{2}(E)\;.
$$
\vskip 3mm

\noindent \textbf{Step 3.} Now we conjugate $C_{1}$ by
$A_{0,2}^{(0,0)}(E)$ to find
\begin{eqnarray*}
\lefteqn{A_{0,2}^{(0,0)}(E)\,C_{1}\,(A_{0,2}^{(0,0)}(E))^{-1}=}
\\ && R_{(0,0)}\left(
\begin{array}{cccc}
\cos^{2}(\alpha_{1})-\sin^{2}(\alpha_{1}) & 0 & -\frac{2}{\alpha_{1}}\sin(\alpha_{1})\cos(\alpha_{1}) & 0 \\
0 & 0 &0 & 0 \\
-2\alpha_{1}\sin(\alpha_{1})\cos(\alpha_{1}) & 0 & \sin^{2}(\alpha_{1})-\cos^{2}(\alpha_{1}) & 0 \\
0 & 0 & 0 & 0
\end{array} \right)R_{(0,0)}^{-1} \in \mathfrak{A}_{2}(E)
\end{eqnarray*}
by (\ref{algconj}). We can subtract from this a multiple of $C_{1}$,
$(\cos^{2}(\alpha_{1})-\sin^{2}(\alpha_{1}))C_{1}$, and divide the
result by $2\alpha_{1}\sin(\alpha_{1})\cos(\alpha_{1})\neq 0$ to
find
$$
C_{3}:=R_{(0,0)} \left( \begin{array}{cccc}
0 & 0 & \frac{1}{\alpha_{1}} & 0 \\
0 & 0 & 0 & 0 \\
\alpha_{1} & 0 & 0 & 0 \\
0 & 0 & 0 & 0
\end{array} \right)R_{(0,0)}^{-1}\in \mathfrak{A}_{2}(E)\;.
$$
Conjugating $C_{1}$ by $A_{0,2}^{(1,1)}(E)$ and repeating the same
arguments we find
$$
C_{4}:=R_{(0,0)} \left( \begin{array}{cccc}
0 & 0 & \frac{1}{\beta_{1}} & 0 \\
0 & 0 & 0 & 0 \\
\beta_{1} & 0 & 0 & 0 \\
0 & 0 & 0 & 0
\end{array} \right)R_{(0,0)}^{-1}\in \mathfrak{A}_{2}(E)\;.
$$

Conjugating $C_{2}$ in the same way shows that
$$
C_{5}:=R_{(0,0)} \left( \begin{array}{cccc}
0 & 0 & 0 & 0 \\
0 & 0 & 0 & \frac{1}{\alpha_{2}} \\
0 & 0 & 0 & 0 \\
0 & \alpha_{2} & 0 & 0
\end{array} \right)R_{(0,0)}^{-1}\in \mathfrak{A}_{2}(E)
$$
and
$$
C_{6}:=R_{(0,0)} \left( \begin{array}{cccc}
0 & 0 & 0 & 0 \\
0 & 0 & 0 & \frac{1}{\beta_{2}} \\
0 & 0 & 0 & 0 \\
0 & \beta_{2} & 0 & 0
\end{array} \right)R_{(0,0)}^{-1}\in \mathfrak{A}_{2}(E)\;.
$$
\vskip 3mm

\noindent \textbf{Step 4.} As $|\alpha_{1}|\neq |\beta_{1}|$ and
$|\alpha_{2}|\neq |\beta_{2}|$ it is clear that the matrices
$C_{1},\ldots ,C_{6}$ are linearly independent. It follows that
\begin{equation}\label{subspace6}
R_{(0,0)} \left( \begin{array}{cccc}
a & 0 & b & 0 \\
0 & \tilde{a} & 0 & \tilde{b} \\
c & 0 & -a & 0 \\
0 & \tilde{c} & 0 & -\tilde{a}
\end{array} \right)R_{(0,0)}^{-1}\in \mathfrak{A}_{2}(E)
\end{equation}
for all $(a,\tilde{a},b,\tilde{b},c,\tilde{c})\in \R^{6}$. \vskip
3mm

\noindent \textbf{Step 5.} Let $\mathcal{C}_{3}$ be the countable
set of energies $E$ such that
$(2\pi,\sqrt{E-\frac{1+\sqrt{5}}{2}},\sqrt{E-\frac{1-\sqrt{5}}{2}})$
is rationally dependent. Then for $E\in (2,+\infty ) \setminus
\mathcal{C}_{3}$, using the same argument as in (\ref{alpha1G}) for
the powers of $A_{0,2}^{(1,0)}(E)$, we have
$$
M_{1}:=R_{(1,0)} \left( \begin{array}{cccc}
0 & 0 & \frac{1}{\alpha} & 0 \\
0 & 1 & 0 & 0 \\
-\alpha & 0 & 0 & 0 \\
0 & 0 & 0 & 1
\end{array} \right)R_{(1,0)}^{-1} \in G_{\tilde{\mu}_E}
$$
where $\alpha:=\sqrt{E-\frac{1+\sqrt{5}}{2}}$. \vskip 3mm

In addition to $C_{1},\ldots ,C_{6}$, we will find four more
linearly independent elements of $\mathfrak{A}_{1}(E)$ by
conjugating particular matrices of the form (\ref{subspace6}) with
$M_1$. Let $X$ be an arbitrary matrix of the form (\ref{subspace6}).
First we remark that
$$
R_{(1,0)}=R_{(0,0)} \left( \begin{array}{cc}
S_{(0,0)}^{-1} S_{(1,0)} & 0 \\
0 & S_{(0,0)}^{-1} S_{(1,0)}
\end{array} \right)\;.
$$
Then a calculation shows that
\begin{equation}\label{formulaMX}
M_{1}XM_{1}^{-1}=R_{(0,0)}\left( \begin{array}{cc}
B & \frac{1}{\alpha}A \\
-\alpha A & B
\end{array} \right)\left( \begin{array}{cccc}
a & 0 & b & 0 \\
0 & \tilde{a} & 0 & \tilde{b} \\
c & 0 & -a & 0 \\
0 & \tilde{c} & 0 & -\tilde{a}
\end{array} \right) \left( \begin{array}{cc}
B & -\frac{1}{\alpha}A \\
\alpha A & B
\end{array} \right)R_{(0,0)}^{-1}\;,
\end{equation}
where $A=T^{-1}\left( \begin{array}{cc}
1 & 0 \\
0 & 0
\end{array} \right)T$, $B=T^{-1}\left( \begin{array}{cc}
0 & 0 \\
0 & 1
\end{array} \right)T$ and $T=S_{(1,0)} S_{(0,0)}^{-1}$.
\vskip 3mm

To construct our last four elements we will take particular values
for $a,\tilde{a},b,\tilde{b},c,\tilde{c}$. Letting $X_{1}$ be the
special case of $X$ where $c=1,
a=0,\tilde{a}=0,b=0,\tilde{b}=0,\tilde{c}=0$, we get, after tedious
calculations,
$$
C_{7}:=M_{1}X_{1}M_{1}^{-1}=\frac{1}{4(5-\sqrt{5})^{2}}R_{(0,0)}
\left( \begin{array}{cccc}
* & -\frac{2+2\sqrt{5}}{\alpha} & * & * \\
\frac{-22+10\sqrt{5}}{\alpha} & * & -\frac{2+2\sqrt{5}}{\alpha^{2}} & * \\
* & 22-10\sqrt{5} & * & * \\
* & * & * & *
\end{array} \right)R_{(0,0)}^{-1} \in \mathfrak{A}_{2}(E)
$$

Here we only keep track of the four matrix elements which are
crucial for establishing linear independence from $C_{1},\ldots
,C_{6}$ as the corresponding matrix-elements of these matrices all
vanish. Similarly to $X_{1}$ we choose $X_{2}$ such that
$\tilde{c}=1$ and all other parameters are $0$. This gives
$$
C_{8}:=M_{1}X_{2}M_{1}^{-1}=\frac{1}{4(5-\sqrt{5})^{2}}R_{(0,0)}
\left( \begin{array}{cccc}
* & \frac{2+2\sqrt{5}}{\alpha} & * & * \\
\frac{22-10\sqrt{5}}{\alpha} & * & \frac{22-10\sqrt{5}}{\alpha^{2}} & * \\
* & -2-2\sqrt{5} & * & * \\
* & * & * & *
\end{array} \right)R_{(0,0)}^{-1} \in \mathfrak{A}_{2}(E)
$$

Under the assumption that
$(2\pi,\sqrt{E-\frac{1+\sqrt{5}}{2}},\sqrt{E-\frac{1-\sqrt{5}}{2}})$
is rationally independent, as in (\ref{alpha2G}), for the powers of
$A_{0,2}^{(1,0)}(E)$ we can prove that
$$
M_{2}:=R_{(1,0)} \left( \begin{array}{cccc}
1 & 0 & 0 & 0 \\
0 & 0 & 0 & \frac{1}{\beta} \\
0 & 0 & 1 & 0 \\
0 & -\beta & 0 & 0
\end{array} \right)R_{(1,0)}^{-1} \in G_{\tilde{\mu}_E}\;,
$$
where $\beta:=\sqrt{E-\frac{1-\sqrt{5}}{2}})$. Then, as before, we
have
$$
M_{2}XM_{2}^{-1}=R_{(0,0)}\left( \begin{array}{cc}
A & \frac{1}{\beta}B \\
-\beta B & A
\end{array} \right)\left( \begin{array}{cccc}
a & 0 & b & 0 \\
0 & \tilde{a} & 0 & \tilde{b} \\
c & 0 & -a & 0 \\
0 & \tilde{c} & 0 & -\tilde{a}
\end{array} \right) \left( \begin{array}{cc}
A & -\frac{1}{\beta}B \\
\beta B & A
\end{array} \right)R_{(0,0)}^{-1}
$$
with the same $A$ and $B$ as in (\ref{formulaMX}). Let $X_{3}$ be
the special case of $X$ with $b=1$ and all the other parameters
equal to $0$ and, similarly, $X_{4}$ with $\tilde{b}=1$ instead of
$b=1$. This gives
$$
C_{9}:=M_{2}X_{3}M_{2}^{-1}=\frac{1}{4(5-\sqrt{5})^{2}}R_{(0,0)}
\left( \begin{array}{cccc}
* & -\beta \frac{1+\sqrt{5}}{8} & * & * \\
\beta \frac{125-41\sqrt{5}}{8} & * &  \frac{1+\sqrt{5}}{8} & * \\
* & \beta^{2}\frac{125-41\sqrt{5}}{8} & * & * \\
* & * & * & *
\end{array} \right)R_{(0,0)}^{-1} \in \mathfrak{A}_{2}(E)
$$
and
$$
C_{10}:=M_{2}X_{4}M_{2}^{-1}=\frac{1}{4(5-\sqrt{5})^{2}}R_{(0,0)}
\left( \begin{array}{cccc}
* & \beta \frac{95-29\sqrt{5}}{8} & * & * \\
\beta \frac{11-5\sqrt{5}}{8} & * & \frac{-11+5\sqrt{5}}{8} & * \\
* & \beta^{2} \frac{95-29\sqrt{5}}{8} & * & * \\
* & * & * & *
\end{array} \right)R_{(0,0)}^{-1} \in \mathfrak{A}_{2}(E)\;.
$$
\vskip 3mm

\noindent \textbf{Step 6.} As $\alpha \neq \beta$, it can be
verified that for most $E$ the four $\R^{4}$-vectors composed of the
four tracked matrix-elements of $C_{7},C_{8},C_{9}$ and $C_{10}$ are
linearly independent. In fact we have
\begin{eqnarray*} \lefteqn{\left|
\begin{array}{cccc}
-\frac{1}{\alpha}(2+2\sqrt{5}) & \frac{1}{\alpha}(2+2\sqrt{5}) & -\beta \frac{1+\sqrt{5}}{8} & \beta \frac{95-29\sqrt{5}}{8} \\[1mm]
\frac{1}{\alpha}(-22+10\sqrt{5}) & \frac{1}{\alpha}(22-10\sqrt{5}) & \beta \frac{125-41\sqrt{5}}{8} & \beta \frac{11-5\sqrt{5}}{8} \\[1mm]
22-10\sqrt{5} & -2-2\sqrt{5} & \beta^{2} \frac{125-41\sqrt{5}}{8} & \beta^{2} \frac{95-29\sqrt{5}}{8} \\[1mm]
-\frac{1}{\alpha^{2}}(2+2\sqrt{5}) &
\frac{1}{\alpha^{2}}(22-10\sqrt{5}) & \frac{1+\sqrt{5}}{8} &
\frac{-11+5\sqrt{5}}{8}
\end{array} \right|} \\ & = & \frac{2\beta(780-349\sqrt{5}(\alpha+\beta)(121\alpha-13664\sqrt{5}\beta-71805\beta))}{121(4(5-\sqrt{5})^{2})^{4}\alpha^{3}}
\;.
\end{eqnarray*}
The right hand side is algebraic as a function of $E$ and therefore
has a discrete set $\mathcal{C}_4$ of zeros.

\noindent Let $\mathcal{C}=\mathcal{C}_{1} \cup \mathcal{C}_{2}\cup
\mathcal{C}_{3} \cup \mathcal{C}_4$. We fix $E\in (2,+\infty )
\setminus \mathcal{C}$. Then $C_{1},\ldots ,C_{6}$ are linearly
independent and so are $C_{7},\ldots,C_{10}$. As the corresponding
matrix-elements of $C_{1},\ldots ,C_{6}$ all vanish, it follows that
$(C_{1},\ldots,C_{10})$ is linearly independent. Thus
$$10 \leq \dim \mathfrak{A}_{2}(E) \leq \dim \mathrm{Sp}_{2}(\R)=10$$
and therefore $\mathfrak{A}_{2}(E) =\mathfrak{sp}_{2}(\R)$. Then by
connectedness of $\mathrm{Sp}_{2}(\R)$ we have
$\mathrm{Cl_{Z}}(G_{\tilde{\mu}_E})=\mathrm{Sp}_{2}(\R)$. We have
proved the proposition.
\end{proof}

\end{document}